\documentclass[a4paper,10pt,twoside]{cpc-hepnp}

\usepackage{multicol}
\usepackage{graphicx}
\usepackage{booktabs}
\usepackage{amssymb,bm,mathrsfs,bbm,amscd}
\usepackage[tbtags]{amsmath}
\usepackage{lastpage}
\usepackage{CJK}

\graphicspath{{Figures/}} %
\usepackage{color}

\newcommand{\mcnum}[2]{$#1\times 10^{#2}$}

\begin{document}

\fancyhead[c]{\small Chinese Physics C~~~Vol. XX, No. X (2014)
XXXXXX} \fancyfoot[C]{\small 010201-\thepage}

\footnotetext[0]{Received 14 May 2014}

\title{Dipion decays of heavy baryons
\thanks{Supported by National Natural Science Foundation of China (11222547, 11175073, 11035006, 11375240, 11261130311), Ministry of Education of China
(FANEDD (200924), DPFIHE (20090211120029), NCET (NCET-10-0442),
Fundamental Research Funds for Central Universities) }}

\author{%
      Chun Mu$^{1;1)}$\email{muchun563@pku.edu.cn}%
\quad Xiao Wang$^{2,3;2)}$\email{xiaowang2011@lzu.edu.cn}%
\quad Xiao-Lin Chen$^{1,2}$ \quad Xiang
Liu$^{2,3;3)}$\email{xiangliu@lzu.edu.cn} \quad Shi-Lin
Zhu$^{1,4;4)}$\email{zhusl@pku.edu.cn} } \maketitle

\address{%
$^1$ Department of Physics and State Key Laboratory of Nuclear
Physics and Technology
and Center of High Energy Physics, Peking University, Beijing 100871, China\\
$^2$ Research Center for Hadron and CSR Physics, Lanzhou University and Institute of Modern Physics of CAS, Lanzhou 730000, China\\
$^3$ School of Physical Science and Technology, Lanzhou University, Lanzhou 730000, China\\
$^4$ Collaborative Innovation Center of Quantum Matter, Beijing 100871, China\\
}

\begin{abstract}
Compared with the charmed baryons, the bottom baryons are not
very well known, either experimentally or theoretically. In this paper, we
investigate the dipion strong decays of the P-wave and D-wave
excited bottom baryons in the framework of the QPC model. We
also extend the same analysis to the charmed baryons.
\end{abstract}



\footnotetext[0]{\hspace*{-3mm}\raisebox{0.3ex}{$\scriptstyle\copyright$}2013
Chinese Physical Society and the Institute of High Energy Physics of
the Chinese Academy of Sciences and the Institute
of Modern Physics of the Chinese Academy of Sciences and IOP Publishing Ltd}%

\begin{multicols}{2}

\section{Introduction}
In 2013, the LHCb Collaboration reported the observations of two
$\Lambda_b^{*0}$ states with the spin-parity quantum numbers $J^{P}=
1/2 ^{-}$ and $3/2 ^{-}$ separately, which are identified as the
orbitally excited states of $\Lambda_{b}^{0}$ \cite{excited_Lambd}.
These observations not only enrich the bottom baryon family, but
also provide important clues to further theoretical and experimental
investigations of the bottom baryons.

Compared with the rich charmed baryon family, the bottom baryons
remain largely unexplored, either experimentally or theoretically.
Experimentally, in addition to the $\Lambda_{b}^{0}$, the
$\Xi_{b}^{-}$ baryon with the quark content $bsd$ was observed by
the D0 \cite{Abazov:2007am} and CDF \cite{Aaltonen:2007ap}
collaborations in 2007. Later, the D0 and CDF collaborations observed
the doubly-strange $\Omega_{b}^{-}$ baryon \cite{Abazov:2008qm,
Aaltonen:2009ny}. Then, the CDF Collaboration observed the ground
state $\Xi_{b}^{0}(bsu)$ with the beauty-strange content
\cite{Aaltonen:2011wd}, and the CMS Collaboration reported the
corresponding excited state, $\Xi_{b}^{0\ast}$ with $J^{P}=3/2^{+}$
\cite{Chatrchyan:2012ni}. Among the triplets $\Sigma^{\pm 0}_{b}$
with spin $J=1/2 $ and $\Sigma^{\ast \pm,0}_b$ with $J=3/2 $, only
the charged states $\Sigma^{(\ast)\pm}_{b}$ were observed in the
$\Lambda_{b}^{0}\pi^{\pm}$ decay modes
\cite{Aaltonen:2007ar,CDF:2011ac}. Theoretically, the two-body
strong decay width of the charmed baryons was investigated several
year ago in Refs. \cite{Chen:2007,Zhong:2007gp}. However, the
three-body strong decays of the heavy baryons are still unexplored
at present.

\begin{center}
\tabcaption{ \label{Table:n1} The mass and discovery channels of the
bottom baryons. Here, we use "--" to denote the case when no strong
decay channel is observed experimentally. The masses are in unit of
MeV.} \footnotesize
\begin{tabular*}{85mm}{@{\extracolsep{\fill}}cccc}
\toprule States & $I(J^{P})$& Mass & Experiment channel \\ \hline
$\Lambda_{b}^{0\ast} $& $0(\frac{1}{2} ^{-})$ & $5912.0 \pm 0.6 $ &
$\Lambda_{b}^{0}\pi^{+}\pi^{-}$
 \cite{excited_Lambd}\\
$\Lambda_{b}^{0\ast} $& $0(\frac{3}{2} ^{-})$ & $5919.8 \pm 0.6 $ &
$\Lambda_{b}^{0}\pi^{+}\pi^{-}$
 \cite{excited_Lambd}\\
 $\Sigma_{b}^{+} $&$1(\frac{1}{2} ^{+})$ & $5811.3 \pm 1.9$ & $\Lambda_{b}^{0}\pi^{+}$ \cite{Aaltonen:2007ar, CDF:2011ac}\\
 $\Sigma_{b}^{-}$&$1(\frac{3}{2} ^{+})$ & $5815.5 \pm 1.8$ & $\Lambda_{b}^{0}\pi^{-}$ \cite{Aaltonen:2007ar, CDF:2011ac}\\
$\Sigma_{b}^{\ast +}$&$1(\frac{1}{2} ^{+})$ & $5832.1 \pm 1.9$ &
$\Lambda_{b}^{0}\pi^{+}$
    \cite{Aaltonen:2007ar, CDF:2011ac}\\
$\Sigma_{b}^{\ast -}$&$1(\frac{3}{2} ^{+})$ & $5835.1 \pm 1.9$ &
$\Lambda_{b}^{0}\pi^{-}$
    \cite{Aaltonen:2007ar, CDF:2011ac}\\
$\Xi_{b}^{0}$&$\frac{1}{2}(\frac{1}{2} ^{+})$ & $5788 \pm 5$ & --
    \cite{Aaltonen:2011wd}\\
$\Xi_{b}^{-}$&$\frac{1}{2}(\frac{1}{2} ^{+})$ & $5791 \pm 2.2$ & --
    \cite{Abazov:2007am, Aaltonen:2007ap}\\
$\Xi_{b}^{0\ast}$&$\frac{1}{2}(\frac{3}{2} ^{+})$ & $5945 \pm 2.3$ &
$\Xi_{b}^{-}\pi^{+}$
    \cite{Chatrchyan:2012ni}\\
\bottomrule
\end{tabular*}%
\end{center}

The experimental progress on the bottom baryons has stimulated
theorists' extensive interest in studying their properties
\cite{Ebert:2007nw,Liu:2007}. In order to understand the structure
of heavy baryons systematically and provide valuable information for
the further experimental exploration, in this work we shall study the dipion
decays of the excited bottom baryons in this work. The dipion decays
are the typical tree-body decays. We adopt the quark pair creation
(QPC) model. We also extend the same formalism to calculate the
dipion decay width of the charmed baryons. The mass and the
corresponding observed decay channel of the bottom baryons are
collected in Table \ref{Table:n1}, while the three-body dipion
decays of the charmed baryons and their corresponding partial decay
widths are listed in Table \ref{Table:nn1}.

\begin{center}
\tabcaption{\label{Table:nn1} A summary of the experimental
three-body dipion decay widths of the charmed baryons in unit of
MeV. 
}
\footnotesize
\begin{tabular*}{85mm}{@{\extracolsep{\fill}}cccc}
\toprule States & $I(J^{P})$& Decay Width & Experiment channel \\
\hline
    $\Lambda_c(2595)^+$ & $0(\frac{1}{2}^-)$ &  $2.6 \pm 0.6 $ & $\Lambda_c^+\pi^+\pi^-$\cite{lambda-2-pi} \\
    $\Lambda_c(2625)^+$ & $0(\frac{3}{2}^-)$ &  $< 0.97      $ & $\Lambda_c^+\pi^+\pi^-$\cite{lambda-2-pi} \\
    $\Lambda_c(2880)^+$ & $0(\frac{5}{2}^+)$ &  $ 5.8\pm 1.1 $ & $\Lambda_c^+\pi^+\pi^-$ \cite{cleo-2880}\\
    $\Xi_c(2815)^+$ & $\frac{1}{2}(\frac{3}{2}^-)$ &  $ < 3.5 $ & $\Xi_c^+\pi^+\pi^-$ \cite{ki-2-pi} \\
    $\Xi_c(2815)^0$ & $\frac{1}{2}(\frac{3}{2}^-)$ &  $ < 6.5 $ & $\Xi_c^0\pi^+\pi^-$ \cite{ki-2-pi}\\
\bottomrule
\end{tabular*}%
\end{center}

In this work, we calculate a special class of the three-body dipion
strong decays of heavy baryons. That is, a heavy baryon decays into
$2\pi$ plus another heavy baryon, where the quantum number of the
$2\pi$ system is either $I(J^P)=0(0^+)$ or $I(J^P)=1(1^{-})$, which
correspond to the intermediate states $\sigma(600)$ and $\rho(770)$,
respectively. Thus, the main task is to calculate the two-body
strong decays of the heavy baryons, where the final states must
contain $\rho(770)$ or $\sigma(600)$. In the next section, we
illustrate the calculation details.

This paper is organized as follows. After the introduction, we
present the the formalism of the three-body dipion strong decays of
the heavy baryons. In Section {\ref{sec3}}, the numerical results
are given. The last section is the discussion and conclusion.

\section{The three-body dipion decays of the bottom baryons}\label{sec2}

We adopt the same notation for the excited heavy baryons as in Ref.
\cite{Chen:2007}. The heavy baryon contains one heavy quark (charm
or bottom) and two light quarks ($u$, $d$ or $s$), which can be
categorized into either the symmetric $6_F$ or antisymmetric
$\bar{3}_F$ flavor representation. For the S-wave heavy baryon, the
total orbital-flavor-spin wave function is symmetric while its color
wave function is antisymmetric. This fact indicates that the spin of
the two light quarks is either $S=1$ for $6_F$ or S=0 for
$\bar{3}_F$. Thus, the spin-parity quantum numbers of the S-wave
heavy baryons are $J^{P}=\frac{1}{2}^+$ or $\frac{3}{2}^{+}$ for
$6_F$ and $J^{P}=\frac{1}{2}^{+}$ for $\bar{3}_F$. Similarly, we can
discuss the P-wave and D-wave heavy baryons. The detailed notations
of the S-wave, P-wave and D-wave heavy baryons can be found in
Figures 1-3 of Ref. \cite{Chen:2007}.

It is difficult to describe the hadron properties, such as the strong
decay process based on the first principle of QCD. Instead, we have to rely
on different phenomenological models to investigate the properties
of the abundant hadronic states. Among these phenomenological
models, the quark pair creation (QPC) model, which was built by Micu
\cite{Micu:1968mk} and further developed by Yaouanc {\it et al.}
\cite{Le Yaouanc:1972ae,Le Yaouanc:1973xz,Le Yaouanc:1974mr,Le
Yaouanc:1977ux,LeYaouanc:1977gm}, has been extensively applied to
study the Okubo-Zweig-Iizuka-allowed strong decays of
hadrons\cite{121,122,123,124,125,126,127,128,129,1401.1595}.

In the QPC model, a pair of the flavor-singlet and color-singlet
light quarks and antiquarks are created from the vacuum, which has the
vacuum quantum number $J^{PC}=0^{++}$. In the non-relativistic
limit, the transition operator is expressed as
\begin{equation}
\begin{split}
 \textit{T} = &-3\gamma \sum_m \langle 1m;1-m\vert 00\rangle \int
    d^3\textbf{k}_4d^3\textbf{k}_5 \delta^3(\textbf{k}_4+\textbf{k}_5)  \\
    &\times
    \mathcal{Y}_{1}^{m}(\frac{\textbf{k}_4-\textbf{k}_5}{2})\chi^{45}_{1,-m}
    \varphi^{45}_{0}\omega_{0}^{45}b^{\dagger}_{4i}(\textbf{k}_4)d^{\dagger}_{5j}(\textbf{k}_5),
\end{split}
\end{equation}
where $i$ and $j$ are the color indices of the created
quark-antiquark pair. And
$\varphi^{45}_{0}=(u\bar{u}+d\bar{d}+s\bar{s})/\sqrt{3}$ and
$\omega^{45}_{0}=\delta_{ij}$ denote the flavor and color wave
functions, respectively, while $\chi_{1,-m}^{45}$ is the spin wave
function with the spin angular momentum $(1,-m)$.
$\mathcal{Y}_{\ell}^{m}(\textbf{k})=\vert\textbf{k}\vert^{\ell}
\textit{Y}_{\ell}^{m}(\theta_{k},\phi_{k})$ is the $\ell$-th solid
harmonic polynomial for the momentum-space distribution of the
quark-antiquark pair. The dimensionless parameter $\gamma$ describes
the strength of the quark-antiquark pair creation from the vacuum.

For the convenience of the calculation, one usually takes the mock
hadron states as follows
\begin{align}
&\vert A(n_{A}^{2S_{A}+1}L_{AJ_AM_{J_A}})(\textbf{P}_A)\rangle
\nonumber\\= & \nonumber
    \sqrt{2E_A} \sum_{M_{L_A},M_{S_A}}\langle L_AM_{L_A}S_AM_{S_A}\vert J_AM_{J_A}\rangle \\&\quad \nonumber
    \times\int d^3\textbf{k}_1d^3\textbf{k}_2d^3\textbf{k}_3
    \delta^3(\textbf{k}_1+\textbf{k}_2+\textbf{k}_3-\textbf{P}_A)\\&\quad \nonumber
    \times\psi_{n_AL_AM_{L_A}}(\textbf{k}_1,\textbf{k}_2,\textbf{k}_3)
    \chi^{123}_{S_AM_{S_A}}\varphi_{A}^{123}\omega^{123}_{A}\\&\quad
    \times\vert  q_1(\textbf{k}_1)q_2(\textbf{k}_2)q_3(\textbf{k}_3)\rangle,
\end{align}
\begin{align}
&\vert B(n_{B}^{2S_{B}+1}L_{BJ_BM_{J_B}})(\textbf{P}_B)\rangle
\nonumber\\ \nonumber= &
    \sqrt{2E_B}\sum_{M_{L_B},M_{S_B}} \langle L_BM_{L_B}S_BM_{S_B}\vert J_BM_{J_B}\rangle \\ \nonumber &\quad  \times
    \int d^3\textbf{k}_ad^3\textbf{k}_b\delta^3(\textbf{k}_a+\textbf{k}_b-\textbf{P}_B)
    \psi_{n_BL_BM_{L_B}}(\textbf{k}_a,\textbf{k}_b) \\ &\quad
    \times\chi_{S_BM_{S_B}}^{ab}\varphi^{ab}_{B}\omega^{ab}_{B} \vert
    q_a(\textbf{k}_a)q_b(\textbf{k}_b)\rangle,
\end{align}
both of which satisfy the normalization conditions
\begin{align}
\langle A(\textbf{P}_A)\vert A(\textbf{P}^{\prime}_A) \rangle
            &= 2E_A\delta^3(\textbf{P}_A-\textbf{P}^{\prime}_{A}),\\
\langle B(\textbf{P}_B)\vert B(\textbf{P}^{\prime}_B) \rangle
            &= 2E_B\delta^3(\textbf{P}_B-\textbf{P}^{\prime}_{B}),
\end{align}
where the subscripts 1, 2, 3 denote the quarks of the parent hadron
$A$ and $a$ and $b$ refer to the quark and antiquark within the
meson $B$, respectively. $\textbf{k}_i$ ($i=1,2,3,a,b$) is the
momentum of the quarks or antiquarks within the hadrons. And
$\textbf{P}_A$ and $\textbf{P}_B$ represent the momentum of $A$ and
$B$, respectively. $S_{A(B)}$ and $J_{A(B)}$ denote the total spin
and total angular momentum of the state $A(B)$, respectively. The
S-matrix of decay is defined as
\begin{equation}
 S=\textit{I}-i2\pi\delta(\textit{E}_{f}-\textit{E}_{i})T.
\end{equation}
In the center of mass frame of baryon $A$, $\textbf{P}_{A}=0$ and
$\textbf{P}_B=-\textbf{P}_C$. Finally, we can formulate the decay
amplitude as
\end{multicols}
\begin{table*}[h]
\begin{equation}\begin{split}
\mathcal{M}^{M_{J_A}M_{J_B}M_{J_C}}=&\langle BC\vert \textit{T}\vert A\rangle\\
    =&\gamma \sqrt{8E_AE_BE_C}\sum_{\substack{M_{L_A},M_{S_A}\\M_{L_B},M_{S_B}\\M_{L_C},M_{S_C},m}}
    \langle L_AM_{L_A}S_AM_{S_A}\vert J_AM_{J_A}\rangle \langle L_BM_{L_B}S_BM_{S_B}\vert J_BM_{J_B}\rangle\langle 1m;1-m\vert 00\rangle\\&
    \times\langle L_CM_{L_C}S_CM_{S_C}\vert J_CM_{J_C}\rangle
    \langle \chi^{235}_{S_CM_{S_C}}\chi^{14}_{S_BM_{S_B}}
        \vert \chi^{123}_{S_AM_{S_A}}\chi^{45}_{1-m}\rangle \langle\varphi_{C}^{235}\varphi_{B}^{14}\vert \varphi_{A}^{123} \varphi_{0}^{45}\rangle
    \textit{I}^{M_{L_A},m}_{M_{L_B},M_{L_C}}(\textbf{P})
\end{split}\end{equation}
\end{table*}
\begin{multicols}{2}
\noindent where the spatial integral
$I^{M_{L_A},m}_{M_{L_B},M_{L_C}}({\textbf{p}})$ is defined as
\begin{align}
   & I^{M_{L_A},m}_{M_{L_B},M_{L_C}}({\textbf{p}})\nonumber\\
   =&   \int\!{\rm d}^3{\textbf{k}}_1{\rm d}^3{\textbf{k}}_2{\rm
        d}^3{\textbf{k}}_3{\rm d}^3{\textbf{k}}_4 {\rm d}^3{\textbf{k}}_5
        \delta^3({\bf{k_4+k_5}})\nonumber\\
    &   \times\delta^3({\bf{k_1+k_2+k_{3}-P_{_A}}})\delta^3({\bf{k_1+k_4-P_{_B}}})\nonumber\\
    &   \times\delta^3({\bf{k_2+k_3+k_{5}-P_{_C}}})
        \nonumber\\
    &   \times\psi^*_{n_B L_B M_{L_B}}\!
        ({\textbf{k}}_1,{\textbf{k}}_4)\psi^*_{n_C L_C M_{L_C}}\!
        ({\textbf{k}}_2,{\textbf{k}}_3,{\textbf{k}}_5)\;\nonumber\\
    &   \times
        \psi_{n_A L_A M_{L_A}}\!
        ({\textbf{k}}_1,{\textbf{k}}_2,{\textbf{k}}_3)\;
        \mathcal{Y}^m_1\Big(\frac{\textbf{k}_4-\text{k}_5}{2}\Big). \label{integral}
\end{align}
and $\langle \chi^{235}_{S_CM_{S_C}}\chi^{14}_{S_BM_{S_B}} \vert
\chi^{123}_{S_AM_{S_A}}\chi^{45}_{1-m}\rangle$ and
$\langle\varphi_{C}^{235}\varphi_{B}^{14}\vert \varphi_{A}^{123}
\varphi_{0}^{45}\rangle$ denote the spin and flavor matrix element,
respectively.

In the framework of the QPC model, the decay occurs through the
recombination of the five quarks from the initial heavy baryon and
the quark-antiquark pair created from the vacuum. There are three
ways of recombination, that is,
\begin{align}
\mathcal{A}(q_1,q_2,Q_3)+\mathcal{P}(\bar{q}_4,q_5) \rightarrow
\mathcal{B}(q_2,Q_3,q_5)+
        \mathcal{C}(q_1,\bar{q}_4), \label{com1}\\
\mathcal{A}(q_1,q_2,Q_3)+\mathcal{P}(\bar{q}_4,q_5) \rightarrow
\mathcal{B}(q_1,Q_3,q_5)+
        \mathcal{C}(q_2,\bar{q}_4),\label{com2}\\
\mathcal{A}(q_1,q_2,Q_3)+\mathcal{P}(\bar{q}_4,q_5) \rightarrow
\mathcal{B}(q_1,q_2,q_5)+
        \mathcal{C}(Q_3,\bar{q}_4).\label{com3}
\end{align}
Here, $Q$ denotes the heavy quark ($b$ or $c$) and $q_i$ is the
light quark. When the excited heavy baryon decays into a heavy
baryon plus a light meson, as shown in Eq. \eqref{com1} and Eq.
\eqref{com2}, the decay amplitude is enhanced by a factor of two:
\begin{equation}\begin{split}
&\mathcal{M}^{M_{J_A}M_{J_B}M_{J_C}}\\
&=2\gamma \sqrt{8E_AE_BE_C}\\&
    \quad\times\sum_{\substack{M_{L_A},M_{S_A}\\M_{L_B},M_{S_B}\\M_{L_C},M_{S_C},m}}
    \langle L_AM_{L_A}S_AM_{S_A}\vert J_AM_{J_A}\rangle \\&
    \quad\times\langle L_BM_{L_B}S_BM_{S_B}\vert J_BM_{J_B}\rangle  \\&
    \quad\times
    \langle L_CM_{L_C}S_CM_{S_C}\vert J_CM_{J_C}\rangle \\&
    \quad\times\langle 1m;1-m\vert 00\rangle
    \langle \chi^{235}_{S_CM_{S_C}}\chi^{14}_{S_BM_{S_B}}
        \vert \chi^{123}_{S_AM_{S_A}}\chi^{45}_{1-m}\rangle\\&
    \quad\times\langle\varphi_{C}^{235}\varphi_{B}^{14}\vert \varphi_{A}^{123} \varphi_{0}^{45}\rangle
    \textit{I}^{M_{L_A},m}_{M_{L_B},M_{L_C}}(\textbf{P}).
\end{split}\end{equation}
However, in the strong decays of $\Xi_{b,c}$ or when the heavy
baryon decays into a heavy meson plus a light baryon, only one way
of arrangement is allowed. Hence this pre-factor two disappears.

The decay width of the process $A\rightarrow B+C$ is
\begin{equation}\label{GAM:n1}
\Gamma =\pi^{2}\frac{\vert p
\vert}{M_A^2}\frac{s}{2J_A+1}\vert\mathcal{M}^{M_{J_A}M_{J_B}M_{J_C}}\vert^{2},
\end{equation}
where $\vert p \vert$ is the outgoing momentum of daughter baryon in
the parent's center mass frame. And $s=1/(1+\delta_{BC})$ is a
statistical factor if $B$ and $C$ are identical particles.

The above formalism is only applicable to the two-body strong
decays. We need to modify the above formalism in order to calculate
the three-body dipion strong decay widths. We assume that the outgoing
meson $B$ is a resonant state of $(\pi\pi)^{I=0}_{l=0}$ or
$(\pi\pi)^{I=1}_{l=1}$. Now the decay width in Eq. \eqref{GAM:n1} is
a function of the mass of the outgoing ``meson'' state $B$. That is,
the original two-body decay width becomes $\Gamma(m_B)$. As a
resonant state of $(\pi\pi)^{I=0}_{l=0}$ or $(\pi\pi)^{I=1}_{l=1}$,
its mass satisfies the Breit-Wigner distribution. We need to
convolute the $\Gamma(m_B)$ with the Breit-Wigner distribution to
get the physical three-pion decay width of the initial particle $A$:
\begin{align}\label{equ_phase}
     \Gamma_{phy}=\int_{2m_{\pi}}^{M_A-M_C}\varrho(m_B)\Gamma(m_B)dm_B,
\end{align}
where $\Gamma(m_B)$ is the decay width of mesons given in the Eq.
\eqref{GAM:n1}. $\varrho(m_B)$ is the Breit-Wigner mass distribution
of the $(\pi\pi)^{I=0}_{l=0}$ or $(\pi\pi)^{I=1}_{l=1}$ resonance
\begin{align}
    \varrho(m_B)=\frac{\Gamma^\prime}{2\pi[(m_B-m_{cen})^2+\Gamma^{\prime 2}/4]},
\end{align}
where $m_{cen}$ and $\Gamma^\prime$ are the mass and decay width of
the $\sigma(600)$ or $\rho(770)$ resonances respectively.

\section{Numerical results}\label{sec3}

The dipion decay widths of the heavy baryons from the QPC model
involve several parameters: the strength of the quark pair creation
from the vacuum $\gamma$, the $R$ value in the harmonic oscillator
wave function of the meson, and the $\alpha_{\rho,\lambda}$ in the
baryon wave functions. There are two kinds of values for $\gamma$
\cite{1401.1595,129,Godfrey,Chen:2007}. We follow the convention of
Ref. \cite{Godfrey} and take $\gamma= 13.4$, which is considered as
a universal parameter in the $^3P_0$ model. The $R$ value of
$\sigma(600)$ mesons is $3.486$ GeV$^{-1}$ \cite{parameter-R1} while
$R=3.571$ GeV$^{-1}$ for the $\rho(770)$ meson \cite{parameter-R1}.
For the proton and $\Lambda$ $\alpha_{\rho}=\alpha_{\lambda}=0.5$
GeV \cite{128}. For the S-wave heavy baryons, the parameters
$\alpha_{\rho}$ and $\alpha_{\lambda}$ in the harmonic oscillator
wave functions can be fixed to reproduce the mass splitting through
the contact term in the potential model \cite{potential}. Their
values are $\alpha_{\rho}=0.6$ GeV and $\alpha_{\lambda}=0.6$ GeV.
For the P-wave and D-wave heavy baryons, $\alpha_{\rho}$ and
$\alpha_{\lambda}$ are expected to lie in the range $0.5\sim 0.7$
GeV. In the following, our numerical results are obtained with the
typical values $\alpha_{\rho}=\alpha_{\lambda}=0.6$ GeV.

For the three-body dipion strong decays, we also need the mass and
width of $(\pi\pi)_{l=0}^{I=0}$ or $(\pi\pi)_{l=1}^{I=1}$,
corresponding to $\sigma(600)$ or $\rho(770)$, which are given in
Table \ref{mass}.
\begin{center}
\tabcaption{The resonance parameters of $\sigma(600)$ and
 $\rho(770)$ corresponding to
 $(\pi\pi)_{l=0}^{I=0}$ and $(\pi\pi)_{l=1}^{I=1}$, respectively.}\label{mass}
\footnotesize
\begin{tabular*}{80mm}{c@{\extracolsep{\fill}}ccccc}
    \toprule Particle & Mass (MeV) & Width (MeV) & $R(\text{GeV}^{-1})$\\\hline
    $\sigma(600)$   & 400 & 400    & 3.486 \cite{parameter-R1}  \\
    $\rho(770)$     & 770 & 149    & 3.571 \cite{parameter-R1}  \\
\bottomrule
\end{tabular*}
\end{center}
With the above preparation, we present the results of the dipion
decay width of of the P-wave and D-wave excited heavy baryons.

\subsection{$1P$ states}

As shown in Tables \ref{Table:n1} and \ref{Table:nn1}, only the
P-wave excited $\Lambda_Q$ ($Q=b,c$) states have so far been observed in the
three-body dipion strong decay channel $\Lambda_Q\pi^+\pi^-$ up to
now, all of which have a small phase space. Here, we present the
theoretical predictions of the three-body dipion strong decays of
all the P-wave excited heavy baryons via the QPC model.

For $\Lambda_{Q}$ ($Q=b,c$), the tiny phase space leads to a very
small strong decay width, which is given in Table
\ref{Tab:Lambda-2625}. Since the $(\pi\pi)_{l=0}^{I=0}$ pair arises
from $\sigma(600)$, the dipion decay width of the heavy baryons
depends on the value of the mass and decay width of $\sigma(600)$.
We present the dependence of the strong decay width of
$\Lambda_{c}(2625)\to \Lambda_c\pi^+\pi^-$ on the width and mass of
$\sigma(600)$ in Table \ref{Tab:Lambda-2625}. The decay width of
$\Lambda_{c}(2625)\to \Lambda_c\pi^+\pi^-$ depends on the width of
$\sigma(600)$ strongly with $m_{\sigma}=400$ MeV. When increasing
$m_{\sigma}$, this dependence becomes weaker.

\begin{center}
\tabcaption{The strong decay width (in unit of keV) of
$\Lambda_{Q}^{\ast}\to \Lambda_Q\pi^+\pi^-$(Q=b,c) when taking
different values of the width of $\sigma(600)$. Here, we fix the
mass of $\sigma(600)$ as 400 MeV.\label{Tab:Lambda-2625}}
\footnotesize
\begin{tabular*}{80mm}{c@{\extracolsep{\fill}}ccccc}
\toprule
&\multicolumn{4}{c}{{$\Gamma^\prime_\sigma$ (GeV)}}\\
{States}            & 0.4   & 0.5   & 0.6   & 0.7   \\\hline
$\Lambda_{c}^\ast(2595)$ & 4.6   & 4     & 3.5   & 3.1   \\
$\Lambda_{c}^\ast(2625)$ & 30.5  & 26.4  & 23.1  & 20.3  \\\hline
$\Lambda_{b}^\ast(5912)$ & 0.77  & 0.68  & 0.6   & 0.53  \\
$\Lambda_{b}^\ast(5920)$ & 2.4   & 2.1   & 1.8   & 1.6   \\
\bottomrule
\end{tabular*}
\end{center}

\begin{center}
\includegraphics[scale=0.67]{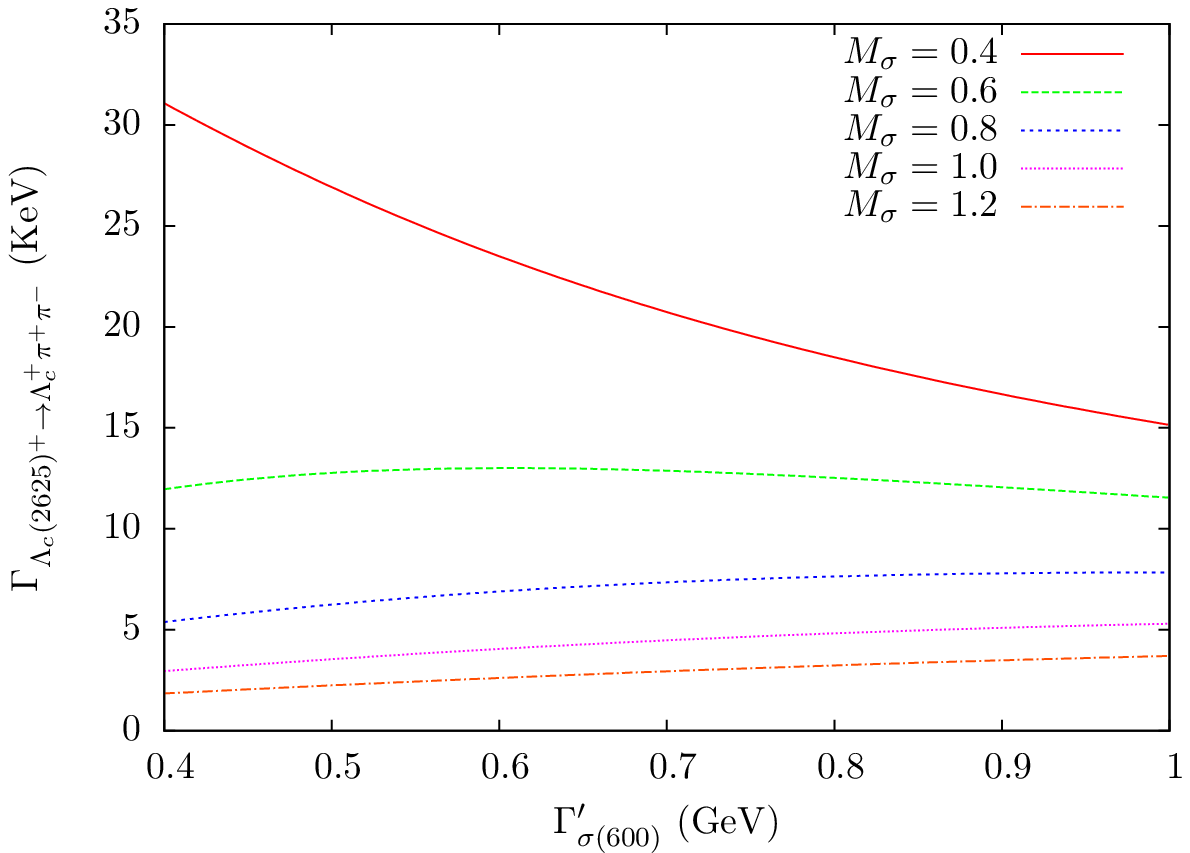}
\figcaption{The variation of the decay width of
$\Lambda_{c}(2625)\to \Lambda_c\pi^+\pi^-$ with the resonance
parameters of $\sigma(600)$\label{lambda-c} }
\end{center}

\begin{center}
\tabcaption{The dipion decay widths of the P-wave excited
$\Xi_{c}(2790)$ and $\Xi_{c}(2815)$ (in unit of
MeV).\label{Tab:P_Ki_c_RS}} \footnotesize
\begin{tabular*}{84mm}{c@{\extracolsep{\fill}}ccccc}
\toprule States  & Assignments  & $I(J^P)$ & $
         \Xi_c(\pi\pi)_{l=1}^{I=1} $ & $ \Xi_c(\pi\pi)_{l=0}^{I=0} $ \\\hline
$\Xi_{c}(2790)$ & $ \Xi_{c1} $ & $\frac{1}{2}(\frac{1}{2} ^-)$ & 0.003 & 0.001     \\
$\Xi_{c}(2815)$ & $ \Xi_{c1} $ & $\frac{1}{2}(\frac{3}{2} ^-)$ & 0.007 & 0.002     \\
\bottomrule
\end{tabular*}
\end{center}

\begin{center}
\tabcaption{The dipion strong decay width of $\Sigma_c(2800)$ with
different assignments (in unit of MeV).\label{Tab:2800-RS}}
\footnotesize
\begin{tabular*}{84mm}{c@{\extracolsep{\fill}}ccccc}
\toprule Assignments & $ \Lambda_c(\pi\pi)_{l=1}^{I=1} $
            & $ \Sigma_c (\pi\pi)_{l=1}^{I=1} $
         & $ \Sigma_c (\pi\pi)_{l=0}^{I=0} $   \\\hline  %
$\Sigma_{c0}(\frac{1}{2} ^-)$             & 0     & 0.038            & 0.003                \\
$\Sigma_{c1}(\frac{1}{2} ^-)$             & 2.6   & \mcnum{6.1}{-4}     & 0.002                \\
$\Sigma_{c1}(\frac{3}{2} ^-)$             & 2.6   & \mcnum{3.1}{-4}     & \mcnum{4.5}{-4}         \\
$\Sigma_{c2}(\frac{3}{2} ^-)$             & 0.078 & 0.38             & 0.11                 \\
$\Sigma_{c2}(\frac{5}{2} ^-)$             & 0.078 & \mcnum{1.2}{-4}     & 0                    \\
$\tilde{\Sigma}_{c1}(\frac{1}{2} ^-)$     & 4.1   & 0.45             & 0.093                \\
$\tilde{\Sigma}_{c1}(\frac{3}{2} ^-)$     & 4.1   & 0.11             & 0.023                \\
\bottomrule
\end{tabular*}
\end{center}

\begin{center}
\tabcaption{The strong decay width of the P-wave excited $\Xi_b$
with different assignments (in unit of MeV).\label{Tab:P_Ki_b_RS}}
\footnotesize
\begin{tabular*}{84mm}{c@{\extracolsep{\fill}}cccc}
\toprule
 States& $ \Xi_b(\pi\pi)_{l=1}^{I=1} $
            & $ \Xi_b^\prime (\pi\pi)_{l=1}^{I=1} $
            & $ \Xi_b (\pi\pi)_{l=0}^{I=0} $
    \\\hline  
$\Xi_{b1}(\frac{1}{2} ^-)$                    & 0.005 & 0     & 0.001  \\
$\Xi_{b1}(\frac{3}{2} ^-)$                    & 0.007 & 0     & 0.002  \\
$\Xi_{b0}^\prime(\frac{1}{2} ^-)$             & 0     & 0.001 & 0      \\
$\Xi_{b1}^\prime(\frac{1}{2} ^-)$             & 0.09  & \mcnum{8}{-7}     & 0.061  \\
$\Xi_{b1}^\prime(\frac{3}{2} ^-)$             & 0.089 & \mcnum{3.5}{-7}     & 0.059  \\
$\Xi_{b2}^\prime(\frac{3}{2} ^-)$             & \mcnum{7.4}{-4}    & 0     & 0      \\
$\Xi_{b2}^\prime(\frac{5}{2} ^-)$             & \mcnum{8.9}{-4}     & 0     & 0      \\
$\tilde{\Xi}_{b1}^\prime(\frac{1}{2} ^-)$     & 0.014 & 0     & 0.004  \\
$\tilde{\Xi}_{b1}^\prime(\frac{3}{2} ^-)$     & 0.02  & 0     & 0.007  \\
$\tilde{\Xi}_{b0}(\frac{1}{2} ^-)$            & 0     & 0.002 & 0      \\
$\tilde{\Xi}_{b1}(\frac{1}{2} ^-)$            & 0.27  & \mcnum{2.5}{-6}     & 0.18   \\
$\tilde{\Xi}_{b1}(\frac{3}{2} ^-)$            & 0.27  & \mcnum{1.1}{-6}     & 0.18   \\
$\tilde{\Xi}_{b2}(\frac{3}{2} ^-)$            & 0.002 & 0     & 0      \\
$\tilde{\Xi}_{b2}(\frac{5}{2} ^-)$            & 0.002 & 0     & 0      \\
\bottomrule
\end{tabular*}
\end{center}

\begin{center}
\tabcaption{The dipion decay widths of the P-wave excited $\Sigma_b$
states with different assignments (in unit of
MeV).\label{Tab:P_Sigma_b_RS}} \footnotesize
\begin{tabular*}{84mm}{c@{\extracolsep{\fill}}ccccc}
\toprule States  &  $ \Lambda_b(\pi\pi)_{l=1}^{I=1} $
        &  $ \Sigma_b(\pi\pi)_{l=1}^{I=1} $
        &  $ \Sigma_b(\pi\pi)_{l=0}^{I=0} $ \\\hline
$\Sigma_{b0}(\frac{1}{2} ^-)$             & 0     & 0.012 & 0.001\\
$\Sigma_{b1}(\frac{1}{2} ^-)$             & 2.3   & \mcnum{5.4}{-6}     & \mcnum{5}{-5}    \\
$\Sigma_{b1}(\frac{3}{2} ^-)$             & 2.2   & \mcnum{1.7}{-6}     & \mcnum{1}{-6}    \\
$\Sigma_{b2}(\frac{3}{2} ^-)$             & 0.031 & 0     & 0    \\
$\Sigma_{b2}(\frac{5}{2} ^-)$             & 0.036 & 0     & 0    \\
$\tilde{\Sigma}_{b1}(\frac{1}{2} ^-)$     & 3.6   & 0.057 & 0.003\\
$\tilde{\Sigma}_{b1}(\frac{3}{2} ^-)$     & 3.6   & 0.014 & 0    \\
\bottomrule
\end{tabular*}
\end{center}

In Tables \ref{Tab:P_Ki_c_RS}-\ref{Tab:P_Sigma_b_RS}, we list the
dipion strong decay widths of the P-wave excited charmed baryons
$\Xi_c(2790)$, $\Xi_c(2815)$, $\Sigma_c(2800)$ and the P-wave
excited states of $\Xi_b$ and $\Sigma_{b}$. The masses and quantum
numbers of $\Xi_c(2790)$ and $\Xi_c(2815)$ are determined
experimentally. However, at present there exits no experimental data for the
P-wave excited $\Xi_{b}$ and $\Sigma_{b}$ at present. The quantum
number of $\Sigma_{c}(2800)$ remains unknown. So, we present the
three-body dipion decay widths of $\Sigma_{c}(2800)$ with different
assignments and the three-body dipion decay widths of the P-wave
excitations of $\Xi_{b}$ and $\Sigma_{b}$ under different
assignments, where the corresponding mass is taken from Ref.
\cite{Ebert:2007nw} if there is no experimental information.

\subsection{$1D$ states}

Quite a few D-wave excited charmed baryons were observed
experimentally, such as $\Lambda_{c}(2880)$ with
$J^P=\frac{5}{2}^{+}$, $\Lambda_{c}(2940)$ with $J^P$ still
undetermined. Both $\Lambda_{c}(2880)$ and $\Lambda_{c}(2940)$ are
considered as the D-wave excited states of $\Lambda_{c}$. Several
other charmed baryons with undetermined $J^P$ quantum numbers were
also considered as the D-wave orbitally excited states, such as
$\Xi_{c}(2980)$ and $\Xi_{c}(3080)$.

Here, we list the dipion decay width of $\Lambda_{c}(2880)$ with
different assignments in Table \ref{Tab:D_Lambda_2880_RS} and that
of $\Lambda_{c}(2940)$ in Table \ref{Tab:D_Lambda_2940_RS}. The
quantum numbers of the D-wave excited $\Xi_{c}(2980,3080,3055,3123)$
are not clear. We list their three-body decay width with different
assignments of their inner quantum numbers. None of the D-wave excited
$\Sigma_{c}$ states and all the D-wave bottom baryons have been
observed experimentally. So, we just give the their three-body dipion
decay width in different assignments of their inner quantum numbers
with their masses chosen from Ref. \cite{Ebert:2007nw} (see Tables
\ref{Tab:D_Sigma_c_RS}-\ref{Tab:D_Ki_b_RS} for more details).

\begin{center}
\tabcaption{The three-body dipion strong decay width of
$\Lambda_c(2880)$ as the D-wave excited states (in unit of
MeV).\label{Tab:D_Lambda_2880_RS}} \footnotesize
\begin{tabular*}{80mm}{c@{\extracolsep{\fill}}ccccc}
\toprule states & $\Sigma_c(\pi\pi)_{l=1}^{I=1}$
       & $\Sigma_c^{\ast}(\pi\pi)_{l=1}^{I=1}$
       & $\Lambda_c(\pi\pi)_{l=0}^{I=0}$                            \\ \hline
$\Lambda_{c2}(\frac{5}{2}^+) $              & 0.002 & 0.002 & \mcnum{2.4}{-3}     \\
$\hat{\Lambda}_{c2}(\frac{5}{2}^+) $              & 0.02  & 0.018 & 0.002     \\
$\check{\Lambda}_{c2}^1(\frac{5}{2}^+) $    & 0     & 0     & 0     \\
$\check{\Lambda}_{c2}^2(\frac{5}{2}^+) $    & 0.002 & 0.007 & 0.3   \\
$\check{\Lambda}_{c3}^2(\frac{5}{2}^+) $    & 0.12  & 0.006 & 0     \\
\bottomrule
\end{tabular*}
\end{center}

\begin{center}
\tabcaption{The strong dipion decay width of $\Lambda_c(2940)$ as
the D-wave excited states (in unit of
MeV).\label{Tab:D_Lambda_2940_RS}} \footnotesize
\begin{tabular*}{80mm}{c@{\extracolsep{\fill}}cccc}
\toprule states & $ \Sigma_c(\pi\pi)_{l=1}^{I=1}  $
       & $ \Sigma_c^\ast(\pi\pi)_{l=1}^{I=1}  $
       & $ \Lambda_c(\pi\pi)_{l=0}^{I=0}  $ \\ \hline
       $\Lambda_{c2}(\frac{3}{2} ^+)$                    & 0.011 & 0.002 & \mcnum{7}{-3}       \\
       $\Lambda_{c2}(\frac{5}{2} ^+)$                    & 0.003 & 0.005 & \mcnum{1}{-4}       \\
$\hat{\Lambda}_{c2}(\frac{3}{2} ^+)$              & 0.1   & 0.022 & 0.007   \\
$\hat{\Lambda}_{c2}(\frac{5}{2} ^+)$              & 0.03  & 0.045 & 0.007   \\
$\check{\Lambda}_{c0}(\frac{1}{2} ^+)$            & 0     & 0     & 0       \\
$\check{\Lambda}_{c1}(\frac{1}{2} ^+)$            & 0     & 0     & 0       \\
$\check{\Lambda}_{c1}(\frac{3}{2} ^+)$            & 0     & 0     & 0       \\
$\check{\Lambda}_{c2}(\frac{3}{2} ^+)$            & 0     & 0     & 0       \\
$\check{\Lambda}_{c2}(\frac{5}{2} ^+)$            & 0     & 0     & 0       \\
$\check{\Lambda}_{c1}^{1}(\frac{1}{2} ^+)$        & 1.9   & 0.13  & 0       \\
$\check{\Lambda}_{c1}^{1}(\frac{3}{2} ^+)$        & 0.23  & 0.64  & 0       \\
$\check{\Lambda}^{2}_{c1}(\frac{1}{2} ^+)$        & 0.016 & 0.007 & 0       \\
$\check{\Lambda}^{2}_{c1}(\frac{3}{2} ^+)$        & 0.013 & 0.008 & 0       \\
$\check{\Lambda}^{2}_{c2}(\frac{3}{2} ^+)$        & 0.045 & 0.004 & 0.65    \\
$\check{\Lambda}^{2}_{c2}(\frac{5}{2} ^+)$        & 0.004 & 0.017 & 0.65    \\
$\check{\Lambda}^{2}_{c3}(\frac{5}{2} ^+)$        & 0.18  & 0.016 & 0       \\
$\check{\Lambda}^{2}_{c3}(\frac{7}{2} ^+)$        & \mcnum{3}{-4}     & 0.072 & 0       \\
\bottomrule
\end{tabular*}
\end{center}
\end{multicols}

\newpage
\begin{center}
\tabcaption{The strong dipion decay width of the D-wave excited
states of $\Sigma_{c}$ (in unit of MeV).\label{Tab:D_Sigma_c_RS}}
\footnotesize
\begin{tabular*}{125mm}{c@{\extracolsep{\fill}}cccccc}
\toprule Assignments & $ \Lambda_c(\pi\pi)_{l=1}^{I=1} $
            & $ \Sigma_c (\pi\pi)_{l=1}^{I=1} $
            & $ \Sigma^\ast_c (\pi\pi)_{l=1}^{I=1} $
            & $ \Sigma_c (\pi\pi)_{l=0}^{I=0} $
            & $ \Sigma_c^\ast (\pi\pi)_{l=0}^{I=0} $  \\\hline
$\Sigma_{c1}(\frac{1}{2} ^{+})$               & 0.3   & 0.015 & 0.01    & 0    & 0.053    \\
$\Sigma_{c1}(\frac{3}{2} ^{+})$               & 0.27  & 0.013 & 0.01    & 0.072  & 0.022   \\
$\Sigma_{c2}(\frac{3}{2} ^{+})$               & 0.55  & 0.041 & 0.006   & 0.006  & 0.002    \\
$\Sigma_{c2}(\frac{5}{2} ^{+})$               & 0.49  & 0.004 & 0.018   & 0.39   & 0.002    \\
$\Sigma_{c3}(\frac{5}{2} ^{+})$               & 0.016 & 0.083 & 0.008   & 0.028    & 0.001   \\
$\Sigma_{c3}(\frac{7}{2} ^{+})$               & 0.04  & 0     & 0.093   & 0   & 0.03    \\
$\hat{\Sigma}_{c1}(\frac{1}{2} ^{+})$         & 2.7   & 0.14  & 0.096   & 0     & 0.48 \\
$\hat{\Sigma}_{c1}(\frac{3}{2} ^{+})$         & 2.5   & 0.12  & 0.09    & 0.654 & 0.2 \\
$\hat{\Sigma}_{c2}(\frac{3}{2} ^{+})$         & 4.9   & 0.37  & 0.054   & 0.06   & 0.02 \\
$\hat{\Sigma}_{c2}(\frac{5}{2} ^{+})$         & 4.4   & 0.036 & 0.16    & 3.5  & 0.026 \\
$\hat{\Sigma}_{c3}(\frac{5}{2} ^{+})$         & 0.15  & 0.75  & 0.073   & 0.25 & 0.016 \\
$\hat{\Sigma}_{c3}(\frac{7}{2} ^{+})$         & 0.36  & 0.006 & 0.84    & 0     & 0.27 \\
$\check{\Sigma}_{c0}^{0}(\frac{1}{2} ^{+})$   & 41    & 3.9   & 3.2     & 0     & 0     \\
$\check{\Sigma}_{c1}^{1}(\frac{1}{2} ^{+})$   & 0     & 0     & 0       & 0     & 0     \\
$\check{\Sigma}_{c1}^{1}(\frac{3}{2} ^{+})$   & 0     & 0     & 0       & 0     & 0     \\
$\check{\Sigma}_{c2}^{2}(\frac{3}{2} ^{+})$   & 2.3   & 0.37  & 0.11    & 0.35  & 0.11  \\
$\check{\Sigma}_{c2}^{2}(\frac{5}{2} ^{+})$   & 2.1   & 0.1   & 0.19    & 0.13  & 0.14  \\
\bottomrule
\end{tabular*}
\end{center}

\newpage

\begin{center}
\tabcaption{The strong dipion decay width of $\Xi_c(2980)$ as the
D-wave excited states (in unit of MeV).\label{Tab:D_Ki_2980_RS}}
\footnotesize
\begin{tabular*}{125mm}{c@{\extracolsep{\fill}}cccccc}
\toprule
    states & $ \Xi_c(\pi\pi)_{l=1}^{I=1}$
    & $ \Xi_c^\prime (\pi\pi)_{l=1}^{I=1}$
    & $ \Xi_c^{\ast\prime} (\pi\pi)_{l=1}^{I=1}$
    & $ \Xi_c(\pi\pi)_{l=0}^{I=0}$
    & $ \Xi_c^\prime(\pi\pi)_{l=0}^{I=0}$
    & $ \Xi_c^{\ast\prime}(\pi\pi)_{l=0}^{I=0}$\\\hline
    $\Xi_{c2}(\frac{3}{2} ^{+})$                      & 0.001 & \mcnum{2.9}{-4}     & \mcnum{2}{-5}     & \mcnum{9.8}{-6} & \mcnum{1.1}{-4} & \mcnum{4.5}{-6}  \\
$\Xi_{c2}(\frac{5}{2} ^{+})$                      & 0.001 & \mcnum{8}{-5}     & \mcnum{4}{-5}     & \mcnum{9.8}{-6}     & \mcnum{4.9}{-5}  & \mcnum{7}{-6}  \\
$\hat{\Xi}_{c2}(\frac{3}{2} ^{+})$                & 0.016 & 0.002 & \mcnum{1.9}{-4}     & \mcnum{8.8}{-5}     & 0.001 & \mcnum{4}{-5}   \\
$\hat{\Xi}_{c2}(\frac{5}{2} ^{+})$                & 0.016 &\mcnum{7.6}{-4} & \mcnum{3.7}{-4} & \mcnum{8.8}{-5} & 0.004 & \mcnum{6.2}{-5}   \\
$\check{\Xi}_{c0}^{1}(\frac{1}{2} ^{+})$          & 0     & 0     & 0     & 0     & 0     & 0     \\
$\check{\Xi}_{c1}^{1}(\frac{1}{2} ^{+})$          & 0     & 0     & 0     & 0     & 0     & 0     \\
$\check{\Xi}_{c1}^{1}(\frac{3}{2} ^{+})$          & 0     & 0     & 0     & 0     & 0     & 0     \\
$\check{\Xi}_{c2}^{1}(\frac{3}{2} ^{+})$          & 0     & 0     & 0     & 0     & 0     & 0     \\
$\check{\Xi}_{c2}^{1}(\frac{5}{2} ^{+})$          & 0     & 0     & 0     & 0     & 0     & 0     \\
$\check{\Xi}_{c1}^{0}(\frac{1}{2} ^{+})$          & 0.12  & 0.047 & 0.001 & 0     & 0.66  & 0     \\
$\check{\Xi}_{c1}^{0}(\frac{3}{2} ^{+})$          & 0.12  & 0.005 & 0.005 & 0     & 0     & 0.15  \\
$\check{\Xi}_{c1}^{2}(\frac{1}{2} ^{+})$          & 0.008 & \mcnum{4.1}{-4} & \mcnum{5}{-5}& 0& 0     & \mcnum{7.7}{-5}    \\
$\check{\Xi}_{c1}^{2}(\frac{3}{2} ^{+})$          & 0.008 & \mcnum{3}{-4} & \mcnum{5}{-5}& 0& \mcnum{9}{-4}     & \mcnum{3.8}{-5}     \\
$\check{\Xi}_{c2}^{2}(\frac{3}{2} ^{+})$          & 0.016 & 0.001 & \mcnum{3}{-5} & 0.019 & \mcnum{9.2}{-5}     & \mcnum{3.9}{-6}     \\
$\check{\Xi}_{c2}^{2}(\frac{5}{2} ^{+})$          & 0.016 & \mcnum{9}{-5} & \mcnum{1.4}{-4}     & 0.019 &  \mcnum{4.1}{-5}     & \mcnum{6.1}{-6}    \\
$\check{\Xi}_{c3}^{2}(\frac{5}{2} ^{+})$          & \mcnum{2.4}{-4}& 0.004 & \mcnum{6}{-4}& 0     & \mcnum{7.5}{-4}     & \mcnum{8.5}{-6}    \\
$\check{\Xi}_{c3}^{2}(\frac{7}{2} ^{+})$          & \mcnum{2.4}{-4}& \mcnum{2}{-6} & \mcnum{6}{-4}& 0     & 0     & \mcnum{3.8}{-5}     \\
${\Xi}^{\prime}_{c1}(\frac{1}{2} ^{+})$           & 0.001 & \mcnum{6}{-5}     & \mcnum{1}{-5} & 0 & 0  & \mcnum{1.3}{-5}  \\
${\Xi}^{\prime}_{c1}(\frac{3}{2} ^{+})$           & 0.001 & \mcnum{5}{-5}     & \mcnum{1}{-5} & 0   & \mcnum{1.6}{-4}  & \mcnum{6.4}{-6}  \\
${\Xi}^{\prime}_{c2}(\frac{3}{2} ^{+})$           & 0.002 & \mcnum{2}{-4}     & \mcnum{6}{-6} & 0.003   & \mcnum{1.5}{-5}  & \mcnum{6.5}{-7} \\
${\Xi}^{\prime}_{c2}(\frac{5}{2} ^{+})$           & 0.002 & \mcnum{1}{-5}     & \mcnum{2}{-5} & 0.003   & \mcnum{6.8}{-6}  & \mcnum{1}{-6}  \\
${\Xi}^{\prime}_{c3}(\frac{5}{2} ^{+})$           & \mcnum{4}{-5}& \mcnum{8}{-4}& \mcnum{2}{-5}& 0  & \mcnum{1.2}{-4}  & \mcnum{1.4}{-6} \\
${\Xi}^{\prime}_{c3}(\frac{7}{2} ^{+})$           & \mcnum{4}{-5}& \mcnum{4}{-7}& \mcnum{1}{-4}& 0  & 0  & \mcnum{6.4}{-6}  \\
$\hat{\Xi}^{\prime}_{c1}(\frac{1}{2} ^{+})$       & 0.013 & \mcnum{6}{-4} & \mcnum{8}{-5}& 0     & 0     & \mcnum{1}{-4}     \\
$\hat{\Xi}^{\prime}_{c1}(\frac{3}{2} ^{+})$       & 0.013 & \mcnum{4}{-4} & \mcnum{1}{-4}& 0     & 0.001  & \mcnum{5}{-5}     \\
$\hat{\Xi}^{\prime}_{c2}(\frac{3}{2} ^{+})$       & 0.024 & 0.001 & \mcnum{5}{-5} & 0.03 & \mcnum{1.3}{-4}  & \mcnum{5.8}{-6}     \\
$\hat{\Xi}^{\prime}_{c2}(\frac{5}{2} ^{+})$       & 0.024 & \mcnum{1}{-4} & \mcnum{2}{-4} & 0.03 & \mcnum{6}{-5} & \mcnum{9.1}{-6}     \\
$\hat{\Xi}^{\prime}_{c3}(\frac{5}{2} ^{+})$       & \mcnum{4}{-4} & 0.007 & \mcnum{2}{-4}& 0 & 0.001 & \mcnum{1}{-5}    \\
$\hat{\Xi}^{\prime}_{c3}(\frac{7}{2} ^{+})$       & \mcnum{4}{-4} & \mcnum{3}{-6}& \mcnum{9}{-4}& 0     & 0     & \mcnum{5}{-5}     \\
$\check{\Xi}^{\prime 0}_{c0}(\frac{1}{2} ^{+})$   & 0.19  & 0.016 & 0.003 & 2.2   & 0     & 0     \\
$\check{\Xi}^{\prime 1}_{c1}(\frac{1}{2} ^{+})$   & 0     & 0     & 0     & 0     & 0     & 0     \\
$\check{\Xi}^{\prime 1}_{c1}(\frac{3}{2} ^{+})$   & 0     & 0     & 0     & 0     & 0     & 0     \\
$\check{\Xi}^{\prime 2}_{c2}(\frac{3}{2} ^{+})$   & 0.011 & 0.001 & \mcnum{1}{-4}& \mcnum{5}{-5} & \mcnum{6.6}{-4} & \mcnum{2}{-5}     \\
$\check{\Xi}^{\prime 2}_{c2}(\frac{5}{2} ^{+})$   & 0.011 & \mcnum{5}{-4}     & \mcnum{2}{-4}& \mcnum{5.8}{-5} &\mcnum{2.9}{-4 } & \mcnum{4.1}{-5}     \\
\bottomrule
\end{tabular*}
\end{center}

\newpage

\begin{center}
\tabcaption{The strong dipion decay width of $\Xi_c(3055)$ as the
D-wave excited states (in unit of MeV).\label{Tab:D_Ki_3055_RS}}
\footnotesize
\begin{tabular*}{125mm}{c@{\extracolsep{\fill}}ccccccc}
\toprule
    states & $ \Xi_c(\pi\pi)_{l=1}^{I=1}$
    & $ \Xi_c^\prime (\pi\pi)_{l=1}^{I=1}$
    & $ \Xi_c^{\ast\prime} (\pi\pi)_{l=1}^{I=1}$
    & $ \Xi_c(\pi\pi)_{l=0}^{I=0}$
    & $ \Xi_c^\prime(\pi\pi)_{l=0}^{I=0}$
    & $ \Xi_c^{\ast\prime}(\pi\pi)_{l=0}^{I=0}$\\\hline
    $\Xi_{c2}(\frac{3}{2} ^{+})$                  & 0.004 & 0.001&\mcnum{2}{-4} & \mcnum{5.4}{-5} & \mcnum{8.6}{-4}   & \mcnum{1.4}{-4}\\
    $\Xi_{c2}(\frac{5}{2} ^{+})$                  & 0.004 &\mcnum{3}{-4} &\mcnum{5}{-4} & \mcnum{5.4}{-5} & \mcnum{3.8}{-4} & \mcnum{2.2}{-4}   \\
$\hat{\Xi}_{c2}(\frac{3}{2} ^{+})$                & 0.044 & 0.011 & 0.002 & \mcnum{4.9}{-4}& 0.008 & 0.001 \\
$\hat{\Xi}_{c2}(\frac{5}{2} ^{+})$                & 0.044 & 0.003 & 0.004 & \mcnum{4.9}{-4} & 0.003 & 0.002\\
$\check{\Xi}_{c0}^{1}(\frac{1}{2} ^{+})$          & 0     & 0     & 0     & 0     & 0     & 0     \\
$\check{\Xi}_{c1}^{1}(\frac{1}{2} ^{+})$          & 0     & 0     & 0     & 0     & 0     & 0     \\
$\check{\Xi}_{c1}^{1}(\frac{3}{2} ^{+})$          & 0     & 0     & 0     & 0     & 0     & 0     \\
$\check{\Xi}_{c2}^{1}(\frac{3}{2} ^{+})$          & 0     & 0     & 0     & 0     & 0     & 0     \\
$\check{\Xi}_{c2}^{1}(\frac{5}{2} ^{+})$          & 0     & 0     & 0     & 0     & 0     & 0     \\
$\check{\Xi}_{c1}^{0}(\frac{1}{2} ^{+})$          & 0.34  & 0.213 & 0.013 & 0     & 1.7   & 0     \\
$\check{\Xi}_{c1}^{0}(\frac{3}{2} ^{+})$          & 0.34  & 0.025 & 0.063 & 0     & 0     & 0.74  \\
$\check{\Xi}_{c1}^{2}(\frac{1}{2} ^{+})$          & 0.023 & 0.001 &\mcnum{7}{-4} & 0     & 0     & 0.002 \\
$\check{\Xi}_{c1}^{2}(\frac{3}{2} ^{+})$          & 0.023 & 0.001 &\mcnum{8}{-4} & 0     & 0.007 & 0.001 \\
$\check{\Xi}_{c2}^{2}(\frac{3}{2} ^{+})$          & 0.043 & 0.005 &\mcnum{4}{-4} & 0.069 & \mcnum{6}{-4} & \mcnum{1}{-4} \\
$\check{\Xi}_{c2}^{2}(\frac{5}{2} ^{+})$          & 0.043 & \mcnum{4}{-4} & 0.001 & 0.069 & \mcnum{3}{-4}&  \mcnum{2}{-4}\\
$\check{\Xi}_{c3}^{2}(\frac{5}{2} ^{+})$          & 0.001 & 0.02  & 0.001 & 0     & 0.006 & \mcnum{3}{-4}\\
$\check{\Xi}_{c3}^{2}(\frac{7}{2} ^{+})$          & 0.001 & \mcnum{3}{-5}& 0.007 & 0     & 0     & 0.001 \\
${\Xi}^{\prime}_{c1}(\frac{1}{2} ^{+})$           & 0.004 &\mcnum{3}{-4} & \mcnum{1}{-4}& 0 & 0  & \mcnum{4}{-4} \\
${\Xi}^{\prime}_{c1}(\frac{3}{2} ^{+})$           & 0.004 &\mcnum{2.5}{-4} &\mcnum{1.4}{-4} & 0 & 0.001  & \mcnum{2}{-2}  \\
${\Xi}^{\prime}_{c2}(\frac{3}{2} ^{+})$           & 0.007 &\mcnum{8}{-4} &\mcnum{7}{-5} & 0.01   & \mcnum{1}{-4}   & \mcnum{1.9}{-5}  \\
${\Xi}^{\prime}_{c2}(\frac{5}{2} ^{+})$           & 0.007 &\mcnum{7}{-5} &\mcnum{3}{-4} & 0.01  & \mcnum{4.9}{-5}  & \mcnum{3}{-5}  \\
${\Xi}^{\prime}_{c3}(\frac{5}{2} ^{+})$           & \mcnum{2}{-4} & 0.003 & \mcnum{3}{-4}     & 0   & 0.001   &\mcnum{4.6}{-5}  \\
${\Xi}^{\prime}_{c3}(\frac{7}{2} ^{+})$           & \mcnum{2}{-4}& \mcnum{6}{-6}& 0.001 & 0  & 0  & \mcnum{2.1}{-4}  \\
$\hat{\Xi}^{\prime}_{c1}(\frac{1}{2} ^{+})$       & 0.035 & 0.002 & 0.001 & 0     & 0     & 0.004 \\
$\hat{\Xi}^{\prime}_{c1}(\frac{3}{2} ^{+})$       & 0.035 & 0.002 & 0.001 & 0     & 0.012 & 0.002 \\
$\hat{\Xi}^{\prime}_{c2}(\frac{3}{2} ^{+})$       & 0.065 & 0.007 & \mcnum{7}{-4} & 0.1 & 0.001 & \mcnum{1.7}{-4}\\
$\hat{\Xi}^{\prime}_{c2}(\frac{5}{2} ^{+})$       & 0.065 & \mcnum{7}{-4 }& 0.002 & 0.1 & \mcnum{4.5}{-4} &\mcnum{2.7}{-4}  \\
$\hat{\Xi}^{\prime}_{c3}(\frac{5}{2} ^{+})$       & 0.001 & 0.03  & 0.002 & 0     & 0.009 & \mcnum{4}{-4} \\
$\hat{\Xi}^{\prime}_{c3}(\frac{7}{2} ^{+})$       & 0.001 & \mcnum{5}{-5} & 0.01  & 0     & 0     & 0.001 \\
$\check{\Xi}^{\prime 0}_{c0}(\frac{1}{2} ^{+})$   & 0.51  & 0.075 & 0.039 & 3.7   & 0     & 0     \\
$\check{\Xi}^{\prime 1}_{c1}(\frac{1}{2} ^{+})$   & 0     & 0     & 0     & 0     & 0     & 0     \\
$\check{\Xi}^{\prime 1}_{c1}(\frac{3}{2} ^{+})$   & 0     & 0     & 0     & 0     & 0     & 0     \\
$\check{\Xi}^{\prime 2}_{c2}(\frac{3}{2} ^{+})$   & 0.029 & 0.007 & 0.001 &\mcnum{3}{-4}  & 0.005 & \mcnum{8}{-4}\\
$\check{\Xi}^{\prime 2}_{c2}(\frac{5}{2} ^{+})$   & 0.029 & 0.002 & 0.003 &\mcnum{3}{-4}  & 0.002 & 0.001 \\
\bottomrule
\end{tabular*}
\end{center}

\newpage

\begin{center}
\tabcaption{The strong dipion decay width of $\Xi_c(3080)$ as the
D-wave excited states (in unit of MeV).\label{Tab:D_Ki_3080_RS}}
\footnotesize
\begin{tabular*}{125mm}{c@{\extracolsep{\fill}}ccccccc}
\toprule
    states & $ \Xi_c(\pi\pi)_{l=1}^{I=1}$
    & $ \Xi_c^\prime (\pi\pi)_{l=1}^{I=1}$
    & $ \Xi_c^{\ast\prime} (\pi\pi)_{l=1}^{I=1}$
    & $ \Xi_c(\pi\pi)_{l=0}^{I=0}$
    & $ \Xi_c^\prime(\pi\pi)_{l=0}^{I=0}$
    & $ \Xi_c^{\ast\prime}(\pi\pi)_{l=0}^{I=0}$\\ \hline
$\Xi_{c2}(\frac{3}{2} ^{+})$                          & 0.006 & 0.001 & \mcnum{4}{-4} &\mcnum{9}{-5} & 0.001   & \mcnum{3}{-4}     \\
$\Xi_{c2}(\frac{5}{2} ^{+})$                          & 0.006 & \mcnum{6}{-4} & \mcnum{8}{-4} &\mcnum{9}{-5} & \mcnum{6.5}{-4} & \mcnum{4.7}{-4} \\
$\hat{\Xi}_{c2}(\frac{3}{2} ^{+})$                    & 0.06  & 0.017 & 0.004 &\mcnum{8}{-4} & 0.013 & 0.003 \\
$\hat{\Xi}_{c2}(\frac{5}{2} ^{+})$                    & 0.06  & 0.005 & 0.007 &\mcnum{8}{-4} & 0.006 & 0.004 \\
$\check{\Xi}_{c0}^{1}(\frac{1}{2} ^{+})$              & 0     & 0     & 0     & 0     & 0     & 0     \\
$\check{\Xi}_{c1}^{1}(\frac{1}{2} ^{+})$              & 0     & 0     & 0     & 0     & 0     & 0     \\
$\check{\Xi}_{c1}^{1}(\frac{3}{2} ^{+})$              & 0     & 0     & 0     & 0     & 0     & 0     \\
$\check{\Xi}_{c2}^{1}(\frac{3}{2} ^{+})$              & 0     & 0     & 0     & 0     & 0     & 0     \\
$\check{\Xi}_{c2}^{1}(\frac{5}{2} ^{+})$              & 0     & 0     & 0     & 0     & 0     & 0     \\
$\check{\Xi}_{c1}^{0}(\frac{1}{2} ^{+})$              & 0.46  & 0.32  & 0.022 & 0     & 2.120 & 0     \\
$\check{\Xi}_{c1}^{0}(\frac{3}{2} ^{+})$              & 0.46  & 0.037 & 0.11  & 0     & 0     & 1     \\
$\check{\Xi}_{c1}^{2}(\frac{1}{2} ^{+})$              & 0.031 & 0.002 & 0.001 & 0     & 0     & 0.005 \\
$\check{\Xi}_{c1}^{2}(\frac{3}{2} ^{+})$              & 0.031 & 0.002 & 0.001 & 0     & 0.011 & 0.002 \\
$\check{\Xi}_{c2}^{2}(\frac{3}{2} ^{+})$              & 0.058 & 0.007 & \mcnum{8}{-4} & 0.098 & 0.001 & \mcnum{2}{-4} \\
$\check{\Xi}_{c2}^{2}(\frac{5}{2} ^{+})$              & 0.058 & \mcnum{7}{-4} & 0.002 & 0.098 & \mcnum{5}{-4} & \mcnum{4}{-4} \\
$\check{\Xi}_{c3}^{2}(\frac{5}{2} ^{+})$              & 0.002 & 0.03  & 0.002 & 0     & 0.01  & \mcnum{6}{-4} \\
$\check{\Xi}_{c3}^{2}(\frac{7}{2} ^{+})$              & 0.002 & \mcnum{6}{-5} & 0.012 & 0     & 0     & 0.002 \\
${\Xi}^{\prime}_{c1}(\frac{1}{2} ^{+})$               & 0.005 & \mcnum{4}{-4}&\mcnum{2}{-4} & 0 & 0 & \mcnum{8.4}{-4}\\
${\Xi}^{\prime}_{c1}(\frac{3}{2} ^{+})$               & 0.005 & \mcnum{4}{-4}&\mcnum{3}{-4} & 0 & 0.002  & \mcnum{4.2}{-4}   \\
${\Xi}^{\prime}_{c2}(\frac{3}{2} ^{+})$               & 0.009 & 0.001 & \mcnum{1}{-4} & 0.016  & \mcnum{1.8}{-4}   & \mcnum{4}{-5} \\
${\Xi}^{\prime}_{c2}(\frac{5}{2} ^{+})$               & 0.009 & \mcnum{1}{-4}  & \mcnum{5}{-4} & 0.016 & \mcnum{8}{-4} & \mcnum{6}{-5}   \\
${\Xi}^{\prime}_{c3}(\frac{5}{2} ^{+})$               & \mcnum{4}{-4} & 0.005 & \mcnum{4}{-4} & 0  & 0.002  & \mcnum{9.8}{-5}  \\
${\Xi}^{\prime}_{c3}(\frac{7}{2} ^{+})$               & \mcnum{4}{-4} & \mcnum{1}{-5}  & 0.002 & 0 & 0 & \mcnum{4}{-4}  \\
$\hat{\Xi}^{\prime}_{c1}(\frac{1}{2} ^{+})$           & 0.047 & 0.004 & 0.001 & 0     & 0     & 0.007 \\
$\hat{\Xi}^{\prime}_{c1}(\frac{3}{2} ^{+})$           & 0.047 & 0.003 & 0.002 & 0 & 0.018 & 0.004 \\
$\hat{\Xi}^{\prime}_{c2}(\frac{3}{2} ^{+})$           & 0.087 & 0.011 & 0.001 & 0.15  & \mcnum{1.7}{-3} & \mcnum{3.6}{-4} \\
$\hat{\Xi}^{\prime}_{c2}(\frac{5}{2} ^{+})$           & 0.087 & 0.001 & 0.004 & 0.15  & \mcnum{7.3}{-4} & \mcnum{5.7}{-4} \\
$\hat{\Xi}^{\prime}_{c3}(\frac{5}{2} ^{+})$           & 0.003 & 0.045 & 0.004 & 0     & 0.015 & \mcnum{8.8}{-4} \\
$\hat{\Xi}^{\prime}_{c3}(\frac{7}{2} ^{+})$           & 0.003 & \mcnum{1}{-4} & 0.018 & 0     & 0     & \mcnum{3.9}{-3} \\
$\check{\Xi}^{\prime 0}_{c0}(\frac{1}{2} ^{+})$       & 0.69  & 0.11  & 0.068 & 4.3   & 0     & 0     \\
$\check{\Xi}^{\prime 1}_{c1}(\frac{1}{2} ^{+})$       & 0     & 0     & 0     & 0     & 0     & 0     \\
$\check{\Xi}^{\prime 1}_{c1}(\frac{3}{2} ^{+})$       & 0     & 0     & 0     & 0     & 0     & 0     \\
$\check{\Xi}^{\prime 2}_{c2}(\frac{3}{2} ^{+})$       & 0.04  & 0.011 & 0.002 & \mcnum{5}{-4} & 0.008 & 0.001 \\
$\check{\Xi}^{\prime 2}_{c2}(\frac{5}{2} ^{+})$       & 0.04  & 0.003 & 0.005 & \mcnum{5}{-4} & 0.003 & 0.002 \\
\bottomrule
\end{tabular*}
\end{center}

\newpage

\begin{center}
\tabcaption{The strong dipion decay width of $\Xi_c(3123)$ as the
D-wave excited states (in unit of MeV).\label{Tab:D_Ki_3123_RS}}
\footnotesize
\begin{tabular*}{125mm}{c@{\extracolsep{\fill}}ccccccc}
\toprule
    states & $ \Xi_c(\pi\pi)_{l=1}^{I=1}$
    & $ \Xi_c^\prime (\pi\pi)_{l=1}^{I=1}$
    & $ \Xi_c^{\ast\prime} (\pi\pi)_{l=1}^{I=1}$
    & $ \Xi_c(\pi\pi)_{l=0}^{I=0}$
    & $ \Xi_c^\prime(\pi\pi)_{l=0}^{I=0}$
    & $ \Xi_c^{\ast\prime}(\pi\pi)_{l=0}^{I=0}$\\\hline
    $\Xi_{c2}(\frac{3}{2} ^{+})$                  & 0.011 & 0.003 & \mcnum{9}{-4}& \mcnum{2}{-4}&\mcnum{3.2}{-3} & \mcnum{8.7}{-4}   \\
    $\Xi_{c2}(\frac{5}{2} ^{+})$                      & 0.011 & 0.001 & 0.001     & \mcnum{2}{-4}& \mcnum{1.4}{-3}   & \mcnum{1.3}{-3}   \\
    $\hat{\Xi}_{c2}(\frac{3}{2} ^{+})$                & 0.099 & 0.031 & 0.008     & \mcnum{1.7}{-3}& 0.028 & \mcnum{7.8}{-3} \\
$\hat{\Xi}_{c2}(\frac{5}{2} ^{+})$                & 0.099 & 0.009 & 0.016     & \mcnum{1.7}{-3}& 0.012 & 0.012 \\
$\check{\Xi}_{c0}^{1}(\frac{1}{2} ^{+})$          & 0     & 0     & 0      & 0     & 0     & 0     \\
$\check{\Xi}_{c1}^{1}(\frac{1}{2} ^{+})$          & 0     & 0     & 0      & 0     & 0     & 0     \\
$\check{\Xi}_{c1}^{1}(\frac{3}{2} ^{+})$          & 0     & 0     & 0      & 0     & 0     & 0     \\
$\check{\Xi}_{c2}^{1}(\frac{3}{2} ^{+})$          & 0     & 0     & 0      & 0     & 0     & 0     \\
$\check{\Xi}_{c2}^{1}(\frac{5}{2} ^{+})$          & 0     & 0     & 0      & 0     & 0     & 0     \\
$\check{\Xi}_{c1}^{0}(\frac{1}{2} ^{+})$          & 0.76  & 0.59  & 0.05   & 0     & 3     & 0     \\
$\check{\Xi}_{c1}^{0}(\frac{3}{2} ^{+})$          & 0.76  & 0.069 & 0.24   & 0     & 0     & 1.7   \\
$\check{\Xi}_{c1}^{2}(\frac{1}{2} ^{+})$          & 0.051 & 0.004 & 0.003  & 0     & 0     & 0.014 \\
$\check{\Xi}_{c1}^{2}(\frac{3}{2} ^{+})$          & 0.051 & 0.004 & 0.003  & 0     & 0.025 & 0.007 \\
$\check{\Xi}_{c2}^{2}(\frac{3}{2} ^{+})$          & 0.095 & 0.013 & 0.001  & 0.17  & 0.002 & \mcnum{6}{-4} \\
$\check{\Xi}_{c2}^{2}(\frac{5}{2} ^{+})$          & 0.095 & 0.001 & 0.006  & 0.17  & 0.001 & 0.001 \\
$\check{\Xi}_{c3}^{2}(\frac{5}{2} ^{+})$          & 0.004 & 0.055 & 0.006  & 0     & 0.022 & 0.001 \\
$\check{\Xi}_{c3}^{2}(\frac{7}{2} ^{+})$          & 0.004 & \mcnum{2}{-4} & 0.026  & 0     & 0     & 0.007 \\
${\Xi}^{\prime}_{c1}(\frac{1}{2} ^{+})$           & 0.008 & \mcnum{8}{-4} & \mcnum{5}{-4} & 0& 0 & \mcnum{2.4}{-3}   \\
${\Xi}^{\prime}_{c1}(\frac{3}{2} ^{+})$           & 0.008 & \mcnum{7}{-4} & \mcnum{6}{-4} & 0   & \mcnum{4.3}{-3}   & \mcnum{1.2}{-3}     \\
${\Xi}^{\prime}_{c2}(\frac{3}{2} ^{+})$           & 0.015 & 0.002 &\mcnum{3}{-4} & 0.028   & \mcnum{3.8}{-4} & \mcnum{1.1}{-4}  \\
${\Xi}^{\prime}_{c2}(\frac{5}{2} ^{+})$           & 0.015 & \mcnum{3}{-4} & 0.001  & 0.028 & \mcnum{1.7}{-4} & \mcnum{1.7}{-4}   \\
${\Xi}^{\prime}_{c3}(\frac{5}{2} ^{+})$           & \mcnum{8}{-4} & 0.009 & 0.001  & 0   & \mcnum{3.8}{-3}   & \mcnum{2.9}{-4}  \\
${\Xi}^{\prime}_{c3}(\frac{7}{2} ^{+})$           & \mcnum{8}{-4} & \mcnum{3}{-4} & 0.004  & 0   & 0  &\mcnum{1.3}{-3}    \\
$\hat{\Xi}^{\prime}_{c1}(\frac{1}{2} ^{+})$       & 0.076 & 0.007 & 0.004  & 0     & 0     & 0.021 \\
$\hat{\Xi}^{\prime}_{c1}(\frac{3}{2} ^{+})$       & 0.076 & 0.006 & 0.005  & 0     & 0.038 & 0.01 \\
$\hat{\Xi}^{\prime}_{c2}(\frac{3}{2} ^{+})$       & 0.14  & 0.02  & 0.002  & 0.25 & \mcnum{3.4}{-3}& \mcnum{1}{-3}\\
$\hat{\Xi}^{\prime}_{c2}(\frac{5}{2} ^{+})$       & 0.14  & 0.002 & 0.009  & 0.25 & \mcnum{1.5}{-3}& \mcnum{1.5}{-3}\\
$\hat{\Xi}^{\prime}_{c3}(\frac{5}{2} ^{+})$       & 0.007 & 0.082 & 0.009  & 0     & 0.034 & \mcnum{2.6}{-3} \\
$\hat{\Xi}^{\prime}_{c3}(\frac{7}{2} ^{+})$       & 0.007 & \mcnum{3}{-5}  & 0     & 0     & 0     & 0.012 \\
$\check{\Xi}^{\prime 0}_{c0}(\frac{1}{2} ^{+})$   & 1.14  & 0.21  & 0.15   & 5.2   & 0     & 0     \\
$\check{\Xi}^{\prime 1}_{c1}(\frac{1}{2} ^{+})$   & 0     & 0     & 0      & 0     & 0     & 0     \\
$\check{\Xi}^{\prime 1}_{c1}(\frac{3}{2} ^{+})$   & 0     & 0     & 0      & 0     & 0     & 0     \\
$\check{\Xi}^{\prime 2}_{c2}(\frac{3}{2} ^{+})$   & 0.066 & 0.02  & 0.005  & 0.001 & 0.019 & 0.005 \\
$\check{\Xi}^{\prime 2}_{c2}(\frac{5}{2} ^{+})$   & 0.066 & 0.006 & 0.011  & 0.001 & 0.008 & 0.008 \\
\bottomrule
\end{tabular*}
\end{center}

\newpage

\begin{center}
\tabcaption{The strong dipion decay width of the D-wave excited
states of $\Lambda_{b}$ (in unit of MeV).\label{Tab:D_Lambda_b_RS}}
\begin{tabular*}{80mm}{c@{\extracolsep{\fill}}cccc}
\toprule states & $\Sigma_c(\pi\pi)_{l=1}^{I=1}$
       & $\Sigma_c^{\ast}(\pi\pi)_{l=1}^{I=1}$
       & $\Lambda_c(\pi\pi)_{l=0}^{I=0}$                                  \\\hline
$\Lambda_{b2}(\frac{3}{2} ^+)$                    & 0.003 & 0.001 &\mcnum{2.6}{-4} \\
$\Lambda_{b2}(\frac{5}{2} ^+)$                    & 0.001 & 0.003 &\mcnum{3.1}{-4}\\
$\hat{\Lambda}_{b2}(\frac{3}{2} ^+)$              & 0.029 & 0.01  &\mcnum{2.3}{-3}\\
$\hat{\Lambda}_{b2}(\frac{5}{2} ^+)$              & 0.01  & 0.027 &\mcnum{2.8}{-3}\\
$\check{\Lambda}_{b0}(\frac{1}{2} ^+)$            & 0     & 0     &0         \\
$\check{\Lambda}_{b1}(\frac{1}{2} ^+)$            & 0     & 0     &0         \\
$\check{\Lambda}_{b1}(\frac{3}{2} ^+)$            & 0     & 0     &0         \\
$\check{\Lambda}_{b2}(\frac{3}{2} ^+)$            & 0     & 0     &0         \\
$\check{\Lambda}_{b2}(\frac{5}{2} ^+)$            & 0     & 0     &0      \\
$\check{\Lambda}_{b1}^{1}(\frac{1}{2} ^+)$        & 0.52  & 0.061 &0      \\
$\check{\Lambda}_{b1}^{1}(\frac{3}{2} ^+)$        & 0.061 & 0.29  &0      \\
$\check{\Lambda}^{2}_{b1}(\frac{1}{2} ^+)$        & 0.004 & 0.003 &0      \\
$\check{\Lambda}^{2}_{b1}(\frac{3}{2} ^+)$        & 0.003 & 0.004 &0      \\
$\check{\Lambda}^{2}_{b2}(\frac{3}{2} ^+)$        & 0.012 & 0.002 &0.32   \\
$\check{\Lambda}^{2}_{b2}(\frac{5}{2} ^+)$        & 0.001 & 0.01  &0.36   \\
$\check{\Lambda}^{2}_{b3}(\frac{5}{2} ^+)$        & 0.065 & 0.009 &0      \\
$\check{\Lambda}^{2}_{b3}(\frac{7}{2} ^+)$        & \mcnum{3}{-5} & 0.044 &0 \\
\bottomrule
\end{tabular*}
\end{center}

\begin{center}
\tabcaption{The strong dipion decay width of the D-wave excited
states of $\Sigma_{b}$ (in unit of MeV).\label{Tab:D_Sigma_b_RS}}
\footnotesize
\begin{tabular*}{120mm}{c@{\extracolsep{\fill}}cccccc}
\toprule Assignments & $ \Lambda_b(\pi\pi)_{l=1}^{I=1} $
            & $ \Sigma_b (\pi\pi)_{l=1}^{I=1} $
            & $ \Sigma^\ast_b (\pi\pi)_{l=1}^{I=1} $
            & $ \Sigma_b (\pi\pi)_{l=0}^{I=0} $
            & $ \Sigma_b^\ast (\pi\pi)_{l=0}^{I=0} $  \\\hline
$\Sigma_{b1}(\frac{1}{2} ^{+})$               & 0.24  & 0.007 & 0.009 & 0  & 0.045     \\
$\Sigma_{b1}(\frac{3}{2} ^{+})$               & 0.1   & 0.001 & 0.002 & \mcnum{6}{-3}   & \mcnum{3}{-3}  \\
$\Sigma_{b2}(\frac{3}{2} ^{+})$               & 0.4   & 0.018 & 0.004 & \mcnum{2.7}{-3} & \mcnum{1.7}{-3}\\
$\Sigma_{b2}(\frac{5}{2} ^{+})$               & 0.35  & 0.001 & 0.013 & \mcnum{9.4}{-4} & \mcnum{2}{-3}\\
$\Sigma_{b3}(\frac{5}{2} ^{+})$               & 0.013 & 0.035 & 0.006 & 0.009    & \mcnum{1.3}{-3}    \\
$\Sigma_{b3}(\frac{7}{2} ^{+})$               & 0.017 & \mcnum{6}{-5} & 0.038 & 0   & 0.009   \\
$\hat{\Sigma}_{b1}(\frac{1}{2} ^{+})$         & 2.1   & 0.069 & 0.081 & 0     & 0.4    \\
$\hat{\Sigma}_{b1}(\frac{3}{2} ^{+})$         & 0.93  & 0.013 & 0.019 & 0.051 & 0.025  \\
$\hat{\Sigma}_{b2}(\frac{3}{2} ^{+})$         & 3.63  & 0.17  & 0.042 & 0.024 & 0.015  \\
$\hat{\Sigma}_{b2}(\frac{5}{2} ^{+})$         & 3.2   & 0.012 & 0.12  & 0.008 & 0.018  \\
$\hat{\Sigma}_{b3}(\frac{5}{2} ^{+})$         & 0.12  & 0.32  & 0.058 & 0.078 & 0.012  \\
$\hat{\Sigma}_{b3}(\frac{7}{2} ^{+})$         & 0.16  & \mcnum{6}{-4} & 0.35  & 0     & 0.082  \\
$\check{\Sigma}_{b0}^{0}(\frac{1}{2} ^{+})$   & 32    & 1.93  & 2.7   & 0     & 0      \\
$\check{\Sigma}_{b1}^{1}(\frac{1}{2} ^{+})$   & 0     & 0     & 0     & 0     & 0      \\
$\check{\Sigma}_{b1}^{1}(\frac{3}{2} ^{+})$   & 0     & 0     & 0     & 0     & 0      \\
$\check{\Sigma}_{b2}^{2}(\frac{3}{2} ^{+})$   & 0.82  & 0.053 & 0.023 & 0.026 & 0.013  \\
$\check{\Sigma}_{b2}^{2}(\frac{5}{2} ^{+})$   & 1.49  & 0.041 & 0.14  & 0.044 & 0.091  \\
\bottomrule
\end{tabular*}
\end{center}

\newpage
\begin{center}
\tabcaption{The strong dipion decay width of the D-wave excited
states of $\Xi_b$ (in unit of MeV).\label{Tab:D_Ki_b_RS}}
\footnotesize
\begin{tabular*}{125mm}{c@{\extracolsep{\fill}}ccccccc}
\toprule
    states & $ \Xi_b(\pi\pi)_{l=1}^{I=1}$
    & $ \Xi_b^\prime (\pi\pi)_{l=1}^{I=1}$
    & $ \Xi_b^{\ast\prime} (\pi\pi)_{l=1}^{I=1}$
    & $ \Xi_b(\pi\pi)_{l=0}^{I=0}$
    & $ \Xi_b^\prime(\pi\pi)_{l=0}^{I=0}$
    & $ \Xi_b^{\ast\prime}(\pi\pi)_{l=0}^{I=0}$\\\hline
$\Xi_{b2}(\frac{3}{2} ^{+})$                          & 0.005 & \mcnum{6}{-4} & \mcnum{2}{-4} & \mcnum{6.7}{-5} & \mcnum{2.8}{-4} & \mcnum{1.1}{-4}\\
$\Xi_{b2}(\frac{5}{2} ^{+})$                          & 0.005 & \mcnum{2}{-4} & \mcnum{5}{-4} & \mcnum{7.6}{-5} &\mcnum{1.4}{-4}  & \mcnum{2.2}{-4}    \\
$\hat{\Xi}_{b2}(\frac{3}{2} ^{+})$                    & 0.047 & 0.005 & 0.001 & \mcnum{6}{-4} & \mcnum{2.5}{-3} & \mcnum{1}{-3} \\
$\hat{\Xi}_{b2}(\frac{5}{2} ^{+})$                    & 0.051 & 0.001 & 0.004 & \mcnum{6.8}{-4} & \mcnum{1.3}{-3} & \mcnum{2}{-3} \\
$\check{\Xi}_{b0}^{1}(\frac{1}{2} ^{+})$              & 0     & 0     & 0     & 0     & 0     & 0      \\
$\check{\Xi}_{b1}^{1}(\frac{1}{2} ^{+})$              & 0     & 0     & 0     & 0     & 0     & 0      \\
$\check{\Xi}_{b1}^{1}(\frac{3}{2} ^{+})$              & 0     & 0     & 0     & 0     & 0     & 0      \\
$\check{\Xi}_{b2}^{1}(\frac{3}{2} ^{+})$              & 0     & 0     & 0     & 0     & 0     & 0      \\
$\check{\Xi}_{b2}^{1}(\frac{5}{2} ^{+})$              & 0     & 0     & 0     & 0     & 0     & 0      \\
$\check{\Xi}_{b1}^{0}(\frac{1}{2} ^{+})$              & 0.36  & 0.092 & 0.011 & 0     & 2.1   & 0      \\
$\check{\Xi}_{b1}^{0}(\frac{3}{2} ^{+})$              & 0.36  & 0.01  & 0.055 & 0     & 0     & 1.4    \\
$\check{\Xi}_{b1}^{2}(\frac{1}{2} ^{+})$              & 0.024 & \mcnum{8}{-4} & \mcnum{6}{-4} & 0     & 0     & 0.012  \\
$\check{\Xi}_{b1}^{2}(\frac{3}{2} ^{+})$              & 0.024 & \mcnum{6}{-4} & \mcnum{7}{-4} & 0     & 0.008 & 0.004  \\
$\check{\Xi}_{b2}^{2}(\frac{3}{2} ^{+})$              & 0.045 & 0.002 & \mcnum{4}{-4} & 0.17  & \mcnum{8}{-4} & \mcnum{4}{-4} \\
$\check{\Xi}_{b2}^{2}(\frac{5}{2} ^{+})$              & 0.049 & \mcnum{2}{-4} & 0.001 & 0.15  & \mcnum{3}{-4} & \mcnum{6}{-4} \\
$\check{\Xi}_{b3}^{2}(\frac{5}{2} ^{+})$              & 0.001 & 0.01  & 0.001 & 0     & 0.006 & \mcnum{9}{-4} \\
$\check{\Xi}_{b3}^{2}(\frac{7}{2} ^{+})$              & 0.001 & \mcnum{1}{-5} & 0.007 & 0     & 0     & 0.002  \\
${\Xi}^{\prime}_{b1}(\frac{1}{2} ^{+})$               & 0.008 & \mcnum{4}{-4}&\mcnum{4}{-4}& 0 & 0  & \mcnum{2}{-3}   \\
${\Xi}^{\prime}_{b1}(\frac{3}{2} ^{+})$               & 0.007 & \mcnum{3}{-4}&\mcnum{4}{-4}& 0   &\mcnum{1.4}{-3} & \mcnum{7.9}{-4}  \\
${\Xi}^{\prime}_{b2}(\frac{3}{2} ^{+})$               & 0.014 & 0.001 &\mcnum{2}{-4} & 0.028 & \mcnum{1.4}{-4}   & \mcnum{7.9}{-5}  \\
${\Xi}^{\prime}_{b2}(\frac{5}{2} ^{+})$               & 0.013 & \mcnum{8}{-5}&\mcnum{7}{-4} & 0.025 & \mcnum{5}{-5} & \mcnum{9.7}{-5}  \\
${\Xi}^{\prime}_{b3}(\frac{5}{2} ^{+})$               & \mcnum{4}{-4}& 0.002 & \mcnum{4}{-4} & 0 &\mcnum{5.4}{-4}  & \mcnum{7.1}{-5}  \\
${\Xi}^{\prime}_{b3}(\frac{7}{2} ^{+})$               & \mcnum{5}{-4}&\mcnum{4}{-6} & 0.002 & 0 & 0  & \mcnum{4.8}{-4}  \\
$\hat{\Xi}^{\prime}_{b1}(\frac{1}{2} ^{+})$           & 0.076 & 0.003 & 0.003 & 0     & 0     & 0.018  \\
$\hat{\Xi}^{\prime}_{b1}(\frac{3}{2} ^{+})$           & 0.068 & 0.002 & 0.003 & 0     & 0.013 & \mcnum{7.1}{-3}  \\
$\hat{\Xi}^{\prime}_{b2}(\frac{3}{2} ^{+})$           & 0.13  & 0.009 & 0.002 & 0.26  & \mcnum{1.3}{-3} & \mcnum{7.1}{-4} \\
$\hat{\Xi}^{\prime}_{b2}(\frac{5}{2} ^{+})$           & 0.12  & \mcnum{7}{-4} & 0.006 & 0.23  & \mcnum{4.7}{-4} & \mcnum{8.8}{-4} \\
$\hat{\Xi}^{\prime}_{b3}(\frac{5}{2} ^{+})$           & 0.003 & 0.019 & 0.003 & 0     & \mcnum{4.9}{-3} & \mcnum{6}{-4} \\
$\hat{\Xi}^{\prime}_{b3}(\frac{7}{2} ^{+})$           & 0.004 & \mcnum{3}{-5} & 0.019 & 0     & 0     & \mcnum{4}{-3} \\
$\check{\Xi}^{\prime 0}_{b0}(\frac{1}{2} ^{+})$       & 1.14  & 0.104 & 0.13  & 4.1   & 0     & 0      \\
$\check{\Xi}^{\prime 1}_{b1}(\frac{1}{2} ^{+})$       & 0     & 0     & 0     & 0     & 0     & 0      \\
$\check{\Xi}^{\prime 1}_{b1}(\frac{3}{2} ^{+})$       & 0     & 0     & 0     & 0     & 0     & 0      \\
$\check{\Xi}^{\prime 2}_{b2}(\frac{3}{2} ^{+})$       & 0.06  & 0.009 & 0.004 & 0     & 0.001 & 0      \\
$\check{\Xi}^{\prime 2}_{b2}(\frac{5}{2} ^{+})$       & 0.054 & 0.002 & 0.007 & 0     & 0     & 0.001  \\
\bottomrule
\end{tabular*}
\end{center}

\begin{multicols}{2}

\section{Discussion and conclusion}

We have performed a systematic investigation of a special class of
the dipion strong decays of the excited heavy baryons where the two
pions are from the intermediate rho or sigma mesons.

The dipion decay width of the heavy baryons is sensitive to the
value of the strength of the quark-pair creation from the vacuum
since the decay width $\Gamma\propto\gamma^2$. However, the
dependence of the dipion decay width on the value of
$\alpha_{\lambda,\rho}$ and $R$ is weak \cite{Chen:2007}. Some
dipion decay modes are forbidden by symmetry and their dipion decay
widths are listed as zero in the tables.


The P-wave excited state $\Sigma_{c}(2800)$ and most of the D-wave
excited states $\Lambda_{c}(2940)$ and $\Xi_c(2980,3080,3055,3123)$
were reported with their quantum numbers undetermined. A
systematic investigation of their two-body strong decays was
presented in Ref. \cite{Chen:2007}, which may be helpful to the
identification of their quantum numbers. Our present work
investigated the dipion decay width of these particles with
different inner structure assignments. For example, the dipion decay
width of $\Sigma_{c}(2800)$ with different internal structures varies
from 1 keV to several MeV as shown in Table \ref{Tab:2800-RS}, which
provides valuable clues to their underlying structure and quantum
numbers. Unfortunately, none of the dipion decay modes have been
 experimentally observed for these states.

The $\Sigma_c\pi$ mode is the dominant decay mode of the P-wave
excited $\Lambda_{c}$ baryon. In contrast, the $\Sigma_b\pi$ mode is
kinematically forbidden for the P-wave excited $\Lambda_{b}$ baryon.
Moreover, the conservation of the isospin symmetry forbids the
$\Lambda_b \pi$ decay mode. In other words, the dipion decay channel
becomes the dominant mode for the P-wave excited $\Lambda_{b}$ heavy
baryons. Because of the tiny phase space, the dipion strong decay
width of these excited states is very small and is less than 5 keV,
which may be comparable to its electromagnetic decay width. In other
words, these two P-wave $\Lambda_{b}$ baryons are extremely narrow
resonances, which may be the most narrow baryon resonances up to
now.

We notice that the different internal structure of the heavy baryon
leads to very different dipion strong decay widths, even if their
$J^P$ quantum numbers are the same. For example, the dipion strong
decay patterns of the three $J^P={1\over 2}^-$ $\Sigma_c$ states are
very different. In other words, the dipion decay modes are very
useful tools to probe the underlying structure of the excited heavy
baryons. Hopefully, the present work will be helpful to the future
experimental search of the excited heavy baryons, and the assignment
of their quantum numbers and internal structures.

\end{multicols}

\vspace{10mm}

\begin{multicols}{2}




\end{multicols}

\vspace{-1mm} \centerline{\rule{80mm}{0.1pt}} \vspace{2mm}

\begin{multicols}{2}

\end{multicols}

\clearpage

\end{document}